\documentclass{rspublic}
\usepackage{graphicx}
\usepackage{amsmath}
\usepackage{amssymb}
\usepackage{natbib}
\usepackage{epsfig}

\newcommand{\upd}{\mathrm{d}}
\newcommand{\dg}{\mbox{$\Delta\gamma$}}
\newcommand{\dn}{\mbox{$\Delta\nu$}}
\newcommand{\cE}{{\cal E}}

\title{The Liquid Blister Test}

\author[J. Chopin, D. Vella and A. Boudaoud]{Julien Chopin, Dominic Vella and Arezki Boudaoud}
\affiliation{Laboratoire de Physique Statistique de l'ENS, UMR8550 du CNRS\\
24, rue Lhomond, 75231 Paris Cedex 05, France}

\begin{document}

\maketitle

\begin{abstract}{Adhesion; elastic plate theory; elastocapillarity; developable surfaces; buckling}
We consider a thin elastic sheet adhering to a stiff substrate by means of the surface tension of a thin liquid layer. Debonding is initiated by imposing a vertical displacement at the centre of the sheet and leads to the formation of a delaminated region, or `blister'. This experiment reveals that the perimeter of the blister takes one of three different forms depending on the vertical displacement imposed. As this displacement is increased, we observe first circular, then undulating and finally triangular blisters. We obtain theoretical predictions for the observed features of each of these three families of blisters. The theory is built upon the F\"{o}ppl-von K\'{a}rm\'{a}n equations for thin elastic plates and accounts for the surface energy of the liquid. We find good quantitative agreement between our theoretical predictions and experimental results, demonstrating that all three families are governed by different balances between elastic and capillary forces. Our results may bear on micrometric tapered devices and other systems where elastic and adhesive forces are in competition.
\end{abstract}

\section{Introduction}

Adhesion is ubiquitous in a range of industrial applications and biological situations. On the one hand, highly optimized adhesives are used to assemble components or to repair broken objects. In electronics and in coating applications, an elastic film often adheres to a substrate through molecular forces~\cite[see][for example]{xia99}. On the other hand, geckos and insects use a combination of Van der Waals and capillary forces to adhere to substrates and walk upside down~\cite[][]{autumn02,huber05,gorb05}.

While industrial applications generally seek permanent bonding, life usually requires reversible bonding. In both cases the strength of bonding is, therefore, a quantity of considerable interest. The most common measure of the strength of bonding is the work of adhesion $\Delta\gamma$, i.e. the energy per unit area needed to create two new interfaces when separating the two adhering objects. Many of the experimental methods developed to measure the work of adhesion actually rely on some combination of adhesive and elastic forces. For example, in the JKR test~\cite[][]{johnson} the deformation of a soft sphere is used to infer the value of $\Delta\gamma$. Similarly, in the `peel test', the work of adhesion is determined from the force required to peel a bonded film from a rigid substrate~\cite[][]{obreimoff30,kendall75}.

As well as providing convenient techniques for measuring the work of adhesion, the interaction of elastic and surface forces gives rise to a rich variety of physical phenomena. In the peeling geometry, for example, fingering instabilies have been observed for both viscous~\cite[][]{mcewan66} and elastic~\cite[][]{ghatak00,adda06} substrates. Collective effects appear in the capillary-induced adhesion of an ensemble of flexible strips or rods~\cite[][]{bico,kim04,py}. These phenomena belong to a broad class of problems involving both elasticity and capillarity, which is referred to as `elasto-capillarity'.
 
Here, we consider a thin sheet adhering to a stiff substrate. In this situation, debonding can either be initiated from the periphery (as when peeling an adhesive strip) or from a point beneath the sheet. Both of these methods have commonly been used as a way of measuring the adhesive energy of polymer coatings on rigid substrates. In the latter situation, a shaft penetrating the substrate is used to push the sheet from below. This causes the formation of an internal delamination blister and so is commonly referred to in the literature as the blister test \cite[][]{dannenberg61,briscoe91}. While this is a useful test for characterizing the properties of adhesives, it is distinct from the delamination observed in many manufacturing processes, which is often driven by pre-existing stresses within the substrate \cite[][]{gioia}.

The experiment presented in this article is similar to the blister test except that the adhesion is mediated by the surface tension of a liquid --- hence this experiment may be thought of as the `liquid blister test'. More precisely, an elastic disk is bound to a substrate by a very thin liquid layer and quasi-statically loaded from below by means of a central indentor. This experimental setting allows us to go beyond the standard blister test and to unravel previously unreported regimes where the blister is not axisymmetric. We find an instability driven by orthoradial compression in an annulus, but the selection of the wavelength is different from that found in other systems~\cite[][]{mora06,huang07}.  The quasi-static nature of the experiment also distinguishes this instability from the fingering observed when a fluid displaces a more viscous liquid~\cite[][]{saffman58}. At very large displacements, we find that the blister takes a triangular shape. The adhered part of the sheet then forms a conical shape reminiscent of the developable cones observed in crumpled paper~\cite[][]{benamar97,cerda99}.

The paper is organized as follows. In \S 2, we detail the experimental setup and characterize experimentally the main regimes.  These regimes are identified by the shape of the blister edge (circular, undulating or triangular) and depend on the displacement imposed by the indentor. In \S 3, we model the system as a thin elastic plate adhering to a stiff substrate with a constant work of adhesion. Using this model we study the properties of the blisters in each of the three regimes and compare the results to our experimental data. Finally, in \S 4, we summarize our findings and discuss some of their implications.

\section{The experiment}

\subsection{Experimental setup}
\label{sec:ExperimentalSetup}

\begin{figure}[ht]
\centering
\includegraphics[height=4cm]{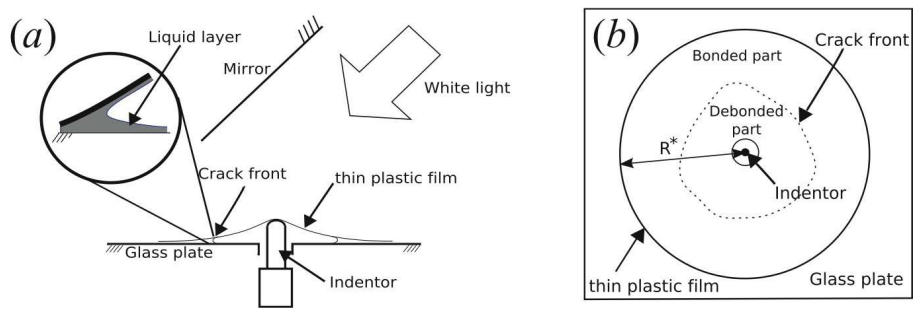}
\caption{Experimental setup. Side (a) and plan (b) views. (a) A thin plastic sheet (see table~\ref{tab:mech_ppty} for properties) of radius $R^*$ adheres to a glass plate ($330\times300\textrm{mm}^2$). The adhesion is mediated by a thin layer of ethanol. A hole of 16mm diameter is drilled in the middle of the plate for an indentor (cylinder of  $6.3$mm in diameter, capped by a hemisphere) that can produce displacements of up to 2cm. (b) When moving the indentor upwards, a delamination blister is nucleated and a crack front propagates (dashed curve).}
\label{fig:schema_manip}
\end{figure}

\begin{table}
\begin{center}
\begin{tabular}{|c|c|c|c|c|c|}
\hline
Material&$E$ (GPa)&$\nu$&$h$ ($\mu$m) &$B$ ($\mu$J)&$2R^*$ (mm)\\
\hline
LDPE& 0.2 & 0.4 & 30; 100 & 0.50 &120; 175; 275 \\
\hline
PP (i)& 2.1 & 0.4 & 40 & 13 &120; 175; 275 \\
\hline
PP (ii)& 1.2 & 0.4 & 100 & 120 &120; 175; 275 \\
\hline
PA & 2.7 & 0.4 & 200 & 2100 &175 \\
\hline
\end{tabular}
\end{center}
\caption{Properties of the sheets used: Young's modulus $E$, Poisson ratio $\nu$ ,  thickness $h$, bending stiffness $B$ and disk diameter $2R^*$. LDPE stands for Low Density PolyEthylene, PP for PolyPropylene, PA for PolyAmide-Nylon6. The values of $E$ were averaged over two orthogonal directions as these materials are anisotropic (anisotropy in the range 10--20\%).}
\label{tab:mech_ppty}
\end{table}

The system, see figure \ref{fig:schema_manip}, consists of a thin elastic sheet adhering to a rigid glass plate (of dimensions $330\times300\textrm{mm}^2$) by means of a thin liquid layer. The experiments reported here were performed with ethanol (surface tension $\gamma = 21$mNm$^{-1}$), but we also used silicon oils with no noticeable effect on the results. Ethanol was used for its ease of cleaning (compared to silicon oils) and the insensitivity of its surface tension to contaminants (compared to water). The plate had a hole of diameter 16mm drilled in its centre, which allows for the equilibration of air pressure between the two sides of the sheet. We use  a micrometric screw positioned beneath the hole as an indentor. Its cylindrical end is capped with a hemisphere and has a diameter of 6.3mm. The screw allows a vertical positioning of the cap to within $1\mu$m over a $2$cm range. The hemispherical cap of the indentor pushes the sheet causing debonding of the sheet and nucleation of the blister. As in the delamination literature, we shall use the term `crack front' to designate the frontier between the bonded and the debonded parts of the sheet. Upon debonding, the sheet and the glass substrate remain wetted by the liquid, as shown schematically in the inset of figure \ref{fig:schema_manip}a. The work of adhesion therefore corresponds to the energy required to create two air/ethanol interfaces of surface tension $\gamma$, and  we have $\Delta\gamma=2\gamma$.

Typically the indentor is moved at a vertical speed of $100\mathrm{\mu ms^{-1}}$. Using typical speeds in the region of $5\mathrm{\mu ms^{-1}}$ we see no observable differences. Likewise, with silicon oils of viscosities different to that of ethanol we do not see any observable difference. We therefore conclude that our system is effectively quasi-static and restrict the present study to equilibrium states.

We used circular sheets made of three types of plastic material: Low Density PolyEthylene (LDPE), PolyPropylene (PP) and PolyAmide (PA), supplied by Goodfellow. Their mechanical and geometrical properties are given in table \ref{tab:mech_ppty}. Typically, the Young's modulus $E$ is of the order of 1GPa, the Poisson ratio $\nu\approx0.4$ and the thickness $h$ is in the range 30--200$\mu$m, resulting in a bending stiffness
\begin{equation}
B\equiv \frac{Eh^3}{12(1-\nu^2)},
\label{bendstiff}
\end{equation} in the range 0.5-2000$\mu$J. The diameter, $2R^*$, of the sheets was in the range 12--27.5cm. The initial volume of ethanol used as adhesive was in the range 20--1000$\mu$l depending on the radius of the disk. This corresponds to a liquid layer thickness in the range 1--50$\mu$m. We found that most results were insensitive to the volume of ethanol, except in an intermediate regime that will be discussed below. We therefore did not systematically measure the quantity of ethanol used. Care was taken when laying the sheet on the substrate, in order to avoid pre-stress and to spread a uniform liquid layer.

To aid visualization, the underside of the glass plate was painted white. The blister was lit uniformly using a halogen lamp. Top views of blisters were then taken using a digital camera, with a mirror inclined at $45^{\circ}$ with respect to the glass plate. A home-made image processing method was used to detect the position of the crack front.

\begin{figure}[t]
\centering
\includegraphics[height=6cm]{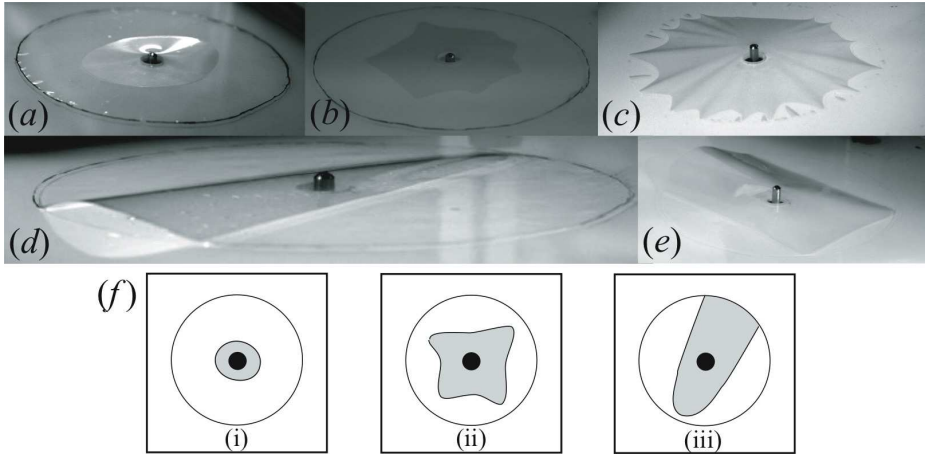}
\caption{The different regimes for the blister shape (sheets of diameter $2R^*=180$mm). (a) A roughly circular crack front at small indentor height ($d=3$mm, LDPE sheet, thickness $h=30\mu$m). (b)  At moderate indentor height ($d = 1.5 - 6$mm),   the crack front becomes unstable and oscillates with a centrimetric wavelength (PP, $h=100\mu$m).  (c)  
The undulation might evolve into a star shape ($d=10$mm, LDPE sheet, thickness $h=30\mu$m). (d) The crack front becomes triangular after the tip of one `finger' has reached the edge ($d=3$mm, PP, $h=100\mu$m). (e) Sometimes the crack front becomes rectangular ($d=3$mm, PP, $h=100\mu$m), instead of triangular. (f) Schematics of the blister as seen from above.}
\label{fig:blisters}
\end{figure}

Figure \ref{fig:blisters} shows images of the blisters as seen from the side. Three main regimes are observed and are shown schematically in figure \ref{fig:blisters}f. At small indentor height $d$, the blister is roughly circular. At moderate $d$, the crack front oscillates. When the thickness of the ethanol layer is small, the undulations develop into a star shape at higher $d$. This undulating regime is the only one which is sensitive to the volume of the adhesive liquid. When $d$ is further increased, one of the fingers of the debonded region reaches the edge of the sheet. The crack front then becomes an open curve, which is generally triangular, but sometimes rectangular. We focus in the following on the more generic regimes, i.e. circular, undulating and triangular blisters.

\subsection{The mean radius of a closed blister}
\label{sec:TheCircularBlister}

\begin{figure}[t]
\centering
\includegraphics[height=6.5cm]{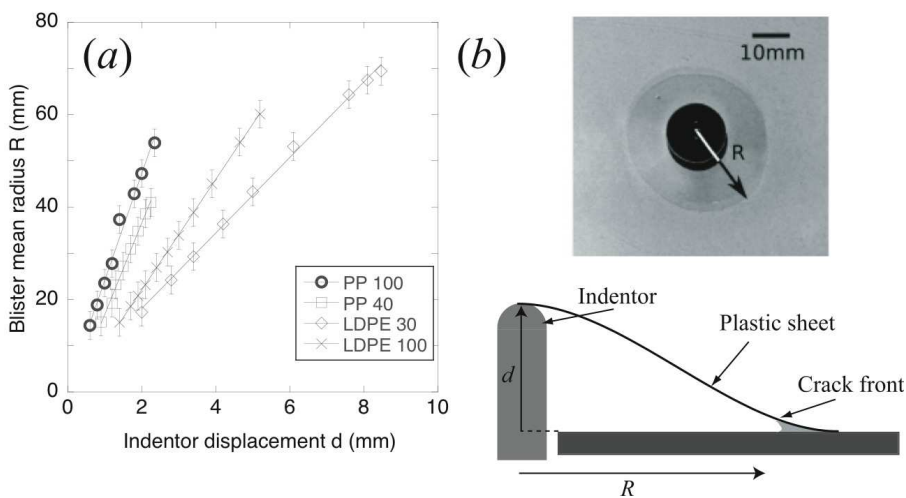}
\caption{(a) Mean radius of a closed blister $R$ vs. indentor height $d$. The symbols (see key) correspond to material type and sheet thickness as detailed in Table~\ref{tab:mech_ppty}; all sheets had  a radius $R^*=175$mm.  (b) Top view and side schematic of a circular blister.}
\label{fig:circ_data}
\end{figure}

The blister is approximately circular at small indentor height $d$ and undulates for larger values of $d$. We measured the mean radius $R$ of the blister as a function of $d$ in both of these regimes. We found $R(d)$ to be linear, with an offset of order $10\mu$m for $d=0$. This offset is due to a small indeterminacy in measuring the displacement when the indentor touches the sheet. Figure \ref{fig:circ_data} therefore shows the curves  $R(d)$ shifted to ensure that they pass through the origin. The linear dependence $R(d)$ also holds approximately when the blister crack front is undulating, and it is insensitive to the radius of the sheet $R^*$ (data not shown). Thus, the slope $R'(d)$ appears to depend only on the thickness of the sheet, on its mechanical properties and possibly on the properties of the liquid. This dependancy will be clarified by the theoretical analysis.

\subsection{The instability}
\label{sec:TheInstability}
\begin{figure}[ht]
\centering
\includegraphics[height=5cm]{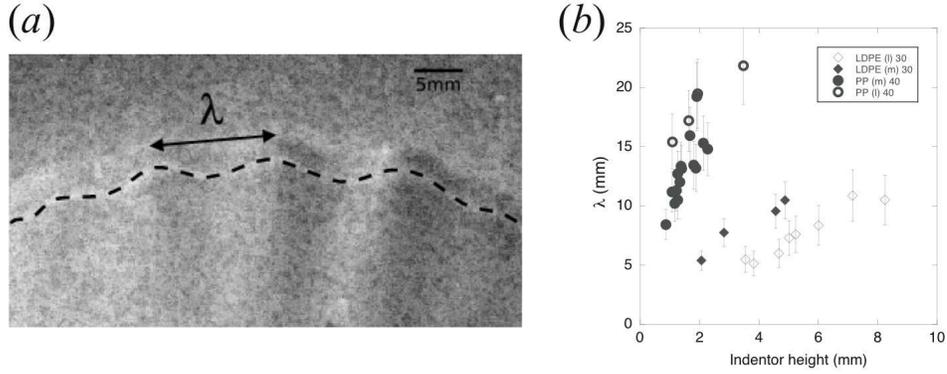}
\caption{The instability: undulations of the blister edge. (a) Top view of a portion of the crack front (LDPE, $h=30\mu$m, $2R^*=$175mm, $d=8$mm). (b) Wavelength $\lambda$ of the undulations as a function of the indentor height $d$. In these experiments, we used  30$\mu$m thick LDPE sheets (diamonds) and 40$\mu$m thick PP (circles). For each material, we used two sheet diameters $2R^*$: 175mm (filled symbols) and 275mm (empty symbols).}
\label{fig:instability}
\end{figure}

It appears that the blister loses axisymmetry for sufficiently large displacements $d>d_c$. The threshold height $d_\mathrm{c}$ at which the circular crack front starts to undulate is highly sensitive to both the thickness of the liquid layer and the preparation of the sheet on the substrate. $d_\mathrm{c}$ tends to decrease for thick fluid layers or when the sheet is not perfectly flat initially. We were not able to determine this threshold $d_\mathrm{c}$ in a reproducible manner. However, we were able to reproducibly measure the wavelengths of the undulations for the thinnest sheets. Figure \ref{fig:instability} shows these experimental results. We note that the wavelength of instability increases with sheet stiffness but seems rather insensitive to the radius of the sheet. Furthermore, the wavelength does not appear to be sensitive to the rate of loading and the undulations persist once the indentor stops moving --- in contrast with the instability of \cite{saffman58}.

\subsection{The triangular blister}
\label{sec:TheTriangularBlister}

\begin{figure}[ht]
\centering
\includegraphics[height=7cm]{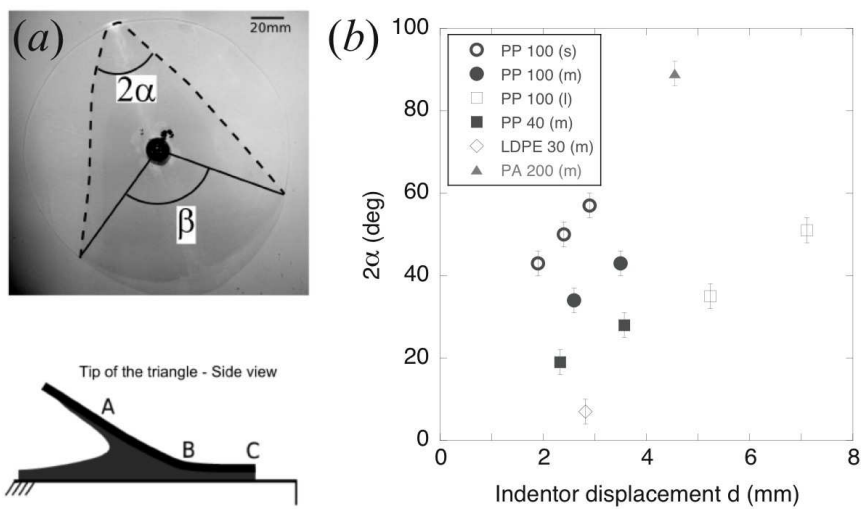}
\caption{The triangular blister. (a) Top: plan view defining the angles $\alpha$ and $\beta$. Bottom: schematic of a horizontal cross section at the tip of the blister, showing the meniscus (A), tip of the blister (B) and edge of the sheet (C). The distance AB is in the millimetric range. (b) Angle of the tip of the triangle $2\alpha$ as a function of the indentor height $d$. The sheets were made of the three plastics (PP, LDPE and PA) with thickness in the range $30-200\mu$m. Three disk diameters were chosen: $120$mm (s), $175$mm (m), $275$mm (l).}
\label{fig:triangular_blister}
\end{figure}

As the undulating crack front approaches the periphery of the sheet, a few undulations grow larger with a finger-like shape, as in figure \ref{fig:blisters}c. With further increases in the indentor height $d$, one of the fingers reaches the edge first. This causes the other fingers to retract, and the crack front becomes an open curve. If the ethanol layer is not too thin ($\gtrsim$20$\mu$m), the edge of the blister relaxes to a triangular shape rounded at one tip (figure \ref{fig:triangular_blister}a). This scenario is made possible by the visible sliding of the sheet along the substrate in the adhering region. Upon further increasing $d$, the angle at the tip of the blister at first increases but the tip disappears at sufficiently large values of $d$. The crack front then consists of two non-intersecting straight lines running between the sheet edges. We do not consider this regime here because of its irreproducibility and instead focus on the range of indentor heights for which the rounded tip (B in figure \ref{fig:triangular_blister}a) is within a few mm of the edge of the sheet (C).

We measured the angle $\beta\simeq4\alpha$, defined in figure \ref{fig:triangular_blister}a, as a function of the indentor displacement $d$. This measurement is well-defined and insensitive to the curved shape of the tip. However, we plot the effective angle at the tip, $2\alpha$, as a function of tip displacement in figure \ref{fig:triangular_blister}b. In contrast to the earlier results for circular and undulating blisters, the radius of the sheet $R^*$ is an important parameter for triangular blisters. As a general trend, the opening angle increases when the bending stiffness $B$ increases or when the radius $R^*$ decreases. We also measured the radius of curvature, $r_c$, of the crack front at the tip as we shall use it in the theoretical section. Typical values of $r_c$ are in the range 1--4cm.

\section{Theory and comparison with experiments}
\label{sec:TheTheory}

The theoretical approach adopted here is to first parametrize the static blister shape in a given regime by a single parameter. For example, we use the blister radius, $R$, as the relevant parameter for circular blisters and the apex angle, $2\alpha$, for triangular blisters. This allows us to calculate the deformation field in the elastic sheet using thin plate theory \cite[][]{mansfield}. Once the deformation field is known, the elastic energy of deformation can then be calculated as a function of this parameter. Adding the surface energy $\dg A$, where $A$ is the area of the blister, we obtain the total energy of deformation ${\cal E}$. The preferred value of the shape parameter, e.g.~$R$ or $2\alpha$, may then be determined by minimizing the energy ${\cal E}$ with respect to variations in this parameter.

\subsection{The circular blister \label{sec:circtheory}}

\subsubsection{Scaling analysis}

We begin by considering a scaling approach for determining the properties of axisymmetric blisters. This approach will be verified in the more detailed analyses that follow, which will also yield the various prefactors in these scalings. An axisymmetric blister of radius $R$ with a vertical displacement $d$ at its centre is subject to a stretching strain $\varepsilon\sim d^2/R^2$ from elementary geometry. The sheet has a stretching stiffness (or spring constant) $Eh$ and so the stretching energy $\cE_e\sim Eh\varepsilon^2 R^2\sim Ehd^4/R^2$. The bending energy $\cE_b$ is proportional to the curvature ($\sim d/R^2$) squared, so $\cE_b\sim B(d/R^2)^2R^2\sim Bd^2/R^2$. Finally, the surface energy required to create the blister is simply the energy required per unit area, $\dg$, multiplied by the blister area and so we have $\cE_s\sim\dg R^2$\footnote{The meniscus near the contact region extends over a region on the order of the capillary length $\ell_c\equiv (\gamma/\rho g)^{1/2}$ where $\rho$ is the liquid density and $g$ the acceleration due to gravity. The gravitational potential energy of the liquid stored in the meniscus is then $\cE_g\sim \rho g \ell_c^2\times R\ell_c\sim \gamma R\ell_c$. Provided that $R\gg \ell_c$ we find that $\cE_s\gg \cE_g$. This is always the case in our experiments and so we neglect this effect in everything that follows.}. The total energy of the system then takes the form
\begin{equation}
{\cal E}\sim\frac{Bd^2}{R^2}+\frac{Ehd^4}{R^2}+\dg R^2.
\label{energy:scal}
\end{equation}

In \eqref{energy:scal}, the blister radius is the only unknown and so we minimize this expression with respect to $R$ to find the blister radius $R$ that is energetically the most favourable. There are two limits that are of particular interest depending on whether $d\ll h$ or $d\gg h$. When $d\ll h$, $\cE_b\gg \cE_e$ and so we find
\begin{equation}
R\sim\left(\frac{B}{\dg}\right)^{1/4}d^{1/2}.
\label{sqrt:scal}
\end{equation} Conversely, when  $d\gg h$, $\cE_e\gg \cE_b$ and we find
\begin{equation}
R\sim\left(\frac{Eh}{\dg}\right)^{1/4}d.
\label{linear:scal}
\end{equation} 

Our experiments are conducted with $d\gg h$ and so we expect to observe the scaling \eqref{linear:scal}. The argument above holds as long as the stretching energy does not vanish. In particular, it applies to axisymmetric configurations which necessarily induce stretching: an axisymmetric sheet takes the form of a cone of revolution, which is impossible for an initially flat sheet unless it is allowed to stretch or is cut. However, as soon as the axisymmetry is broken, configurations with no stretching energy become possible~\cite[they correspond to developable configurations, see][for example]{benamar97,cerda99,cerda05}. For such situations, these scaling analyses reveal that stretching is enormously expensive (relative to bending) and so we expect the system to accommodate deformation via bending. We now move on to determine the prefactor in the scaling relationship \eqref{linear:scal} for an axisymmetric blister.

\subsubsection{General theory}

\begin{figure}
\centering
\includegraphics[height=3.5cm]{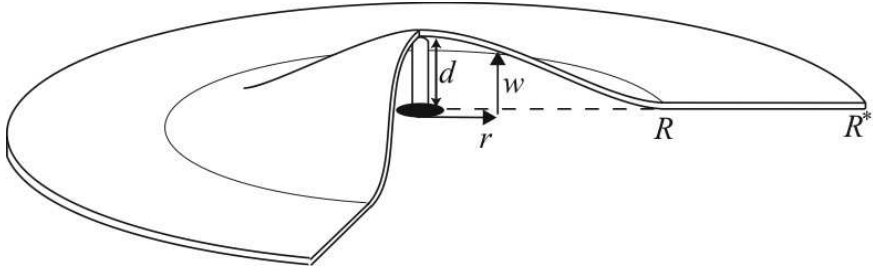}
\caption{Setup and notation for the analysis of an axisymmetric blister.}
\label{blisterdefs}
\end{figure}

A theoretical idealization of an axisymmetric blister is shown in figure \ref{blisterdefs}. In this geometry, the F\"{o}ppl-von K\'{a}rm\'{a}n equations governing the vertical displacement $w$ of the sheet as a function of the radial co-ordinate $r$ become \cite[see][for example]{mansfield}
\begin{equation}
B\frac{\upd}{\upd r}\left\{r\frac{\upd}{\upd r}
\left[\frac{1}{r}\frac{\upd}{\upd r}\bigl(rw'\bigr)\right]\right\}+rq(r)=\frac{\upd}{\upd r}\bigl(\phi'w'\bigr)
\label{dimfvk1}
\end{equation} and
\begin{equation}
\frac{\upd}{\upd r}\left\{r\frac{\upd}{\upd r}
\left[\frac{1}{r}\frac{\upd}{\upd r}\bigl(r\phi'\bigr)\right]\right\}=-\tfrac{1}{2}Eh\frac{\upd}{\upd r}\bigl(w'^2\bigr).
\label{dimfvk2}
\end{equation} Here the bending stiffness $B$ is given in \eqref{bendstiff}, $q$ is a load distribution and  the Airy function $\phi$ is a potential for the stress:
\begin{equation}
\sigma_{rr}=\frac{1}{r}\frac{\upd \phi}{\upd r}\quad\textrm{and}\quad
\sigma_{\theta\theta}=\frac{\upd^2\phi}{\upd r^2}.\label{stressvsairy}
\end{equation}

For a blister of radius $R$, we introduce dimensionless variables
\begin{equation}
\rho\equiv r/R, \quad W\equiv w/R, \quad Q\equiv qR/Eh, \quad \Phi\equiv\phi/EhR^2.
\label{ndimvars}
\end{equation} Within the blistered region ($\rho\leq1$) we assume a point force acts at the origin so that $Q(\rho)=Q\delta(\rho)/2\pi\rho$.

Substituting the dimensionless variables in \eqref{ndimvars} into  \eqref{dimfvk1} and \eqref{dimfvk2} and integrating once with respect to $\rho$ we find that \cite[][]{jensen91}
\begin{equation}
\epsilon\rho\frac{\upd}{\upd \rho}
\left[\frac{1}{\rho}\frac{\upd}{\upd \rho}\bigl(\rho W'\bigr)\right]+\frac{Q}{2\pi}=\Phi'W'
\label{ndfvk1}
\end{equation} and
\begin{equation}
\rho\frac{\upd}{\upd \rho}
\left[\frac{1}{\rho}\frac{\upd}{\upd \rho}\bigl(\rho\Phi'\bigr)\right]=-\tfrac{1}{2}W'^2,
\label{ndfvk2}
\end{equation} where 
\begin{equation}
\epsilon\equiv \frac{B}{EhR^2}=\frac{1}{12(1-\nu^2)}\frac{h^2}{R^2}.
\end{equation} For the values of $h$ and $R$ relevant in our experiments, $\epsilon\ll1$. We therefore expect that the effects of bending may be neglected in comparison with stretching. Furthermore, this assumption should be valid throughout the sheet except in small boundary layers near $\rho=0,1$. 
 
As already discussed, bending dominates stretching for small vertical displacements and there is a transition between bending and stretching behaviours at intermediate displacements, as our earlier scaling analysis shows.  \cite{cotterell97} consider a related problem but their analysis is valid only for $d/h\lesssim10$. The transition between bending and stretching has also been studied by \cite{wan99}, albeit under the assumption that the radial and tangential stresses in the membrane are equal and constant. In \ref{2d:append} we consider the two-dimensional version of this problem. We show there that the scaling \eqref{linear:scal}, with the appropriate prefactor, is correct to within $10\%$ once $d\gtrsim4h$. The assumption that $d\gg h$ is true for all but the very early stages of our axisymmetric experiment and so we shall consider here the `membrane limit' $\epsilon=0$.

Setting $\epsilon=0$ in \eqref{ndfvk1} we find
\begin{equation}
\Phi'W'=\frac{Q}{2\pi},
\label{memlim}
\end{equation} which is to be solved along with \eqref{ndfvk2} and the boundary conditions
\begin{equation}
W(0)=d/R\equiv W_0,\quad W(1)=0,\quad \Phi'(0)=0.
\label{membcs}
\end{equation} (We note that we may arbitrarily set $\Phi(0)$ to zero because only $\Phi'$ enters our equations.) A fourth boundary condition is required to close the system. This boundary condition should be on the horizontal displacement,
\begin{equation}
u_r=\frac{r}{Eh}\bigl(\sigma_{\theta\theta}-\nu\sigma_{rr}\bigr),
\label{rdisp}
\end{equation} at the edge of the blister where $\sigma_{rr}$ and $\sigma_{\theta\theta}$ are the radial and circumferential tensions in the sheet and are given by \eqref{stressvsairy}. However, there are two plausible conditions on $u_r(R)$. If the sheet does not slide outside the blister, then $u_r(R)=0$. If sliding is permitted then we also need to solve the equilibrium equations outside, and we require only continuity of displacements at the blister edge, $u_r(R^+)=u_r(R^-)$. In practice, we might expect some combination of these conditions to be realized. We shall therefore consider both of these possibilities in the following.

\subsubsection{No sliding}

If the horizontal displacement at the blister edge is constrained to be zero, the constitutive equation~\eqref{rdisp} and the definition of the Airy stress function~\eqref{stressvsairy} yield the fourth boundary condition
\begin{equation}
 \Phi''(1)-\nu\Phi'(1)=0.
 \label{bcnoslip}
\end{equation}
We look for solutions of \eqref{ndfvk2} and \eqref{memlim} of the form
$W=W_0(1-\rho^a)$, $\Phi=b\rho^c$, and find that
\begin{equation}
W=W_0(1-\rho^{2/3}), \quad \Phi=\tfrac{3}{16}W_0^2\rho^{4/3}
\label{mansfieldsol}
\end{equation}

This solution is well known and is given, in slightly modified notation, in \cite{mansfield}. However, this solution does not satisfy the boundary condition in \eqref{bcnoslip} unless $\nu=1/3$. For values of $\nu\approx1/3$, a perturbation of \eqref{mansfieldsol} allows us to satisfy this boundary condition, as shown in \ref{app:vneq13}. However, the perturbative solution only changes the final result of the analysis (the predicted blister radius $R$) by less than $3\%$ for $\nu=0.4$, which is a typical value in our experiments. We shall therefore use only the solution in \eqref{mansfieldsol} in the main text.

The deformation field in \eqref{mansfieldsol} gives rise to dimensional stresses (see Eq.~\ref{stressvsairy}),
\begin{equation}
\sigma_{rr}=Eh\frac{W_0^2}{4}\rho^{-2/3},\quad
\sigma_{\theta\theta}=Eh\frac{W_0^2}{12}\rho^{-2/3}\label{sigmaint}
\end{equation} in the sheet.
The stretching energy within the blister is then given by
\begin{eqnarray}
{\cal E}_e&=&\tfrac{1}{2}\int\sigma_{\alpha\beta}\epsilon_{\alpha\beta}~\upd A=\frac{\pi}{Eh}\int_0^Rr(\sigma_{rr}^2-2\nu\sigma_{rr}\sigma_{\theta\theta}+\sigma_{\theta\theta}^2)~\upd r\nonumber\\
&=&\frac{\pi(5-3\nu)}{48}\frac{Ehd^4}{R^2},
\end{eqnarray} where the components of the stress tensor, $\sigma_{\alpha\beta}$, are related to the components of the strain tensor, $\epsilon_{\alpha\beta}$, by the linear elastic constitutive relation \cite[][]{mansfield}.

Combining the stretching energy in the sheet with the change in surface energy associated with a blister of radius $R$, we find that the total energy of the system is
\begin{equation}
{\cal E}=\frac{\pi(5-3\nu)}{48}\frac{Ehd^4}{R^2}+\pi\dg R^2.
\end{equation} Choosing $R$ such that $\partial {\cal E}/\partial R=0$ then gives that
\begin{equation}
R=\tfrac{1}{2}\left\{\bigl(\tfrac{5}{3}-\nu\bigr)\frac{Eh}{\dg}\right\}^{1/4}d.
\label{radiusint}
\end{equation}

\subsubsection{Sliding}

If sliding is allowed beyond the blister, we only require the continuity of horizontal displacement at $\rho=1$,
\begin{equation}
 \Phi''(1^+)-\nu\Phi'(1^+)= \Phi''(1^-)-\nu\Phi'(1^-).
 \label{bcctyslip}
\end{equation} 

Within the blister, it is convenient to introduce
\begin{equation}
\psi(\eta)\equiv \rho \Phi'(\rho),\quad \eta\equiv\rho^2.
\end{equation} 
Along with \eqref{ndfvk1}, this substitution simplifies \eqref{ndfvk2} to
\begin{equation}
\frac{\upd^2 \psi}{\upd \eta^2}=-\frac{Q^2}{32\pi^2} \psi^{-2},
\end{equation} which integrates to
\begin{equation}
\frac{\upd \psi}{\upd \eta}=\frac{Q}{4\pi} \left(\frac{A\psi+1}{\psi}\right)^{1/2}.
\label{ufirstode}
\end{equation}

We then consider the region outside the blister ($\rho>1$) where, by assumption, we have that $W'=0$. However, there is a non-trivial stress function $\Phi$ in this part of the sheet. The most general solution of \eqref{ndfvk2} for $\Phi$ is $\Phi'=a\rho+b/\rho$. Requiring that $\sigma_{rr}(\infty)=0$ gives $a=0$. The constant $b=\psi(1)$ by the continuity of $\sigma_{rr}$ at $\rho=1$, so that 
\begin{equation}
\Phi'=\psi(1)/\rho.\label{airyext}
\end{equation}
The continuity of displacement~\eqref{bcctyslip} then gives $\psi'(1)=0$, which in turn determines $A=-1/\psi(1)$ in \eqref{ufirstode}. Integrating \eqref{ufirstode} analytically, subject to $\Phi'(0)=0$, we find that
\begin{equation}
\frac{\pi}{2}\eta=-\sqrt{\tilde{\psi}\bigl(1-\tilde{\psi}\bigr)}+\tan^{-1}\sqrt{\frac{\tilde{\psi}}{1-\tilde{\psi}}},
\label{sliding1}
\end{equation} 
where $\tilde{\psi}=\psi/\psi(1)$. (We also note that the requirement that $\tilde{\psi}(1)=1$ yields $Q=4\pi\psi(1)^{3/2}$.) This equation gives $\tilde{\psi}$ implicitly in terms of $\eta$. We may also integrate \eqref{ndfvk1} to obtain $W$ as a function of $\tilde{\psi}$, rather than $\eta$. We find
\begin{equation}
W=W_0\left[\tfrac{1}{2}-\tfrac{1}{\pi}\sin^{-1}\bigl(2\tilde{\psi}-1\bigr)\right],\quad \psi(1)=W_0^2/\pi^2.
\label{sliding2}
\end{equation}

The stretching energy for the total deformation of the sheet may be calculated analytically to be
\begin{equation}
{\cal E}_e=\frac{1}{2\pi}\frac{Ehd^4}{R^2},
\end{equation} independently of the Poisson ratio $\nu$. Adding this to the surface energy and minimizing the total energy with respect to variations in $R$, as in the previous section, we find that
\begin{equation}
R=\left(\frac{1}{2\pi^2}\frac{Eh}{\dg}\right)^{1/4}d.
\label{radiusext}
\end{equation}

\begin{figure}
\centering
\includegraphics[height=6cm]{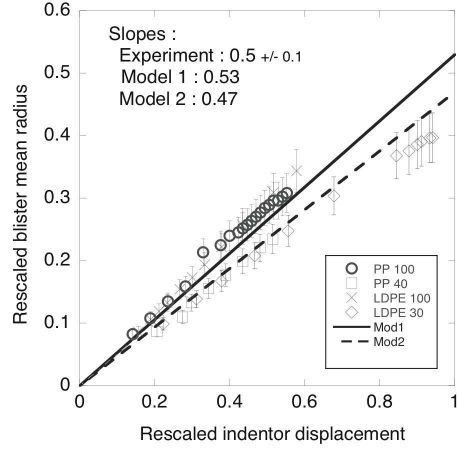}
\caption{The axisymmetric blister. Experimental results for the non-dimensional mean radius of the blister $R / R^*$ as a function of non-dimensional rescaled indentor height $d / [ R^* ( {Eh}/{\Delta \gamma} )^{1/4} ]$. Here $E$ is Young's modulus, $\Delta \gamma$ is twice the surface tension of ethanol, $h$ is the thickness of the sheet and $R^*$ its radius. The theoretical line predicted by model 1 (no sliding outside the blister) has a slope of $0.53$ (solid line), while model 2 (allowing the outside to slide) gives a slope of $0.47$ (dashed line).}
	\label{fig:rescaled_data}
\end{figure}

\subsubsection{Experimental results}
\label{sec:ExperimentalResult}

We first note that equations \eqref{linear:scal},~\eqref{radiusint} and~\eqref{radiusext} all reproduce the linear dependence of the blister radius $R$ on the indentor displacement $d$ observed in experiments. These models also predict that the constant of proportionality is itself proportional to $s\equiv({Eh}/{\Delta \gamma})^{{1}/{4}}$. To go further, we rescale the experimental data of figure \ref{fig:circ_data} using the prefactor $s$. This rescaling allows a collapse of the data as shown in figure \ref{fig:rescaled_data}. The slope of the experimental data is $0.5\pm0.1$, which lies between the slopes $(5/3-\nu)^{1/4}/2\approx0.53$ and $(2\pi^2)^{-1/4}\approx0.47$ given by model 1 (no sliding outside the blister) and model 2  (sliding allowed), respectively. The vertical scatter in the rescaled data are most likely due to the relatively large errors in the value of Young's modulus (10-20\%) caused by the inhomogeneity and anisotropy of the sheet. Although we could not visualize sliding in circular blisters, we believe that it is more likely to occur when the contact is lubricated by a relatively thick ethanol layer; solid friction would prevent sliding if the ethanol layer is very thin. This observation has implications for the instability of the circular blister as discussed in the next section. 

The scaling in \eqref{radiusint} and \eqref{radiusext} has been observed previously in the conventional blister test \cite[][]{briscoe91} with a theoretical analysis presented by \cite{wan95}. However, we note that the coefficient there is slightly different since the analysis of \cite{wan95} explicitly assumes that the beam deflection takes the form of a cone $W=W_0(1-\rho)$. Here we have determined both $W$ and $\Phi$ from the fully coupled problem.

\subsection{Instability\label{sec:Instability}}

\subsubsection{Theory}

In the absence of sliding, we are unable to propose any instability mechanism. However, when sliding is allowed, we find that the azimuthal stress, as resulting from \eqref{airyext},
\begin{equation}
\sigma_{\theta\theta}=-Eh/\pi^2\; (d/r)^2<0,
\label{sigmatt}
\end{equation} 
is compressive outside the blister. (Note that there is always azimuthal traction inside, see e.g. Eq.~\ref{sigmaint}.) This compressive stress could lead to a buckling instability that we analyse here. 

We suppose that this instability gives rise to a vertical displacement of size $H$ with wavelength $\lambda$, which we assume decays over some radius $\tilde{R}$. The bending energy of these fluctuations should be of size $B H^2 \tilde{R}^2/\lambda^4$. A compressive energy within the sheet is released by going out of the plane. The perturbation to the stretching energy has typical size $-|\sigma_{\theta\theta}| H^2\tilde{R}^2 /\lambda^2 $. Finally the undulation of the crack front incurs an energy penalty. The amplitude of the contact line oscillations is  $H/W_0$ where $W_0$ is the slope of the blister, and we find an energy penalty $\dg (H/W_0)^2$ coming from the perturbation to the area of the blister. Adding these three energy terms, we find that
\begin{equation}
{\cal E}\sim H^2\tilde{R}^2\left(\frac{B}{\lambda^4}-\frac{|\sigma_{\theta\theta}|}{\lambda^2}+\frac{\dg}{W_0^2\tilde{R}^2}\right).
\end{equation}

We see that if the term in brackets is positive then increasing the perturbation amplitude $H$ increases the total energy and so is discouraged: the system is stable. Conversely if this term is negative, energy may be released by increasing the perturbation amplitude: the system is unstable. For the system to remain stable we therefore require that
\begin{equation}
|\sigma_{\theta\theta}|\lesssim\frac{B}{\lambda^2}+\frac{\dg \lambda^2}{W_0^2\tilde{R}^2}\equiv f(\lambda).
\label{sigmabd}
\end{equation} 
The minimum value of the function $f(\lambda)$ is attained with
\begin{equation}
\lambda=\lambda_c\sim \left(\frac{BW_0^2\tilde{R}^2}{\dg}\right)^{1/4}\sim 
h \left(\frac{Eh}{\dg}\right)^{1/4}\left(\frac{d}{h}\right)^{1/2}\left(\frac{\tilde{R}}{R}\right)^{1/2}.\label{lambda}
\end{equation}

The equation \eqref{sigmabd} then requires that $|\sigma_{\theta\theta}|\lesssim \min f$ for stability, i.e
\begin{equation}
|\sigma_{\theta\theta}|\lesssim\left(\frac{B\dg}{W_0^2\tilde{R}^2}\right)^{1/2}.
\end{equation}
 Using the expression for $\sigma_{\theta\theta}$ in \eqref{sigmatt} evaluated at $r=R$ (where the compression is largest) we may write the condition of stability as
\begin{equation}
{d}/{R}\lesssim {h}/{\tilde{R}}.\label{condstab}
\end{equation}

This scenario involves fluctuations of the sheet on both sides of the crack front. The compressive stress $\sigma_{\theta\theta}$ outside the blister is comparable to the stress inside as long as $r\sim R$ \eqref{sigmatt}, suggesting that the driving mechanism for the instability acts on a scale $\tilde{R}\simeq R$. The condition \eqref{condstab} then implies that the system is unstable as soon as $d\gtrsim h$.
 
\subsubsection{Experimental results}

\begin{figure}
\centering
\includegraphics[width=7.5cm]{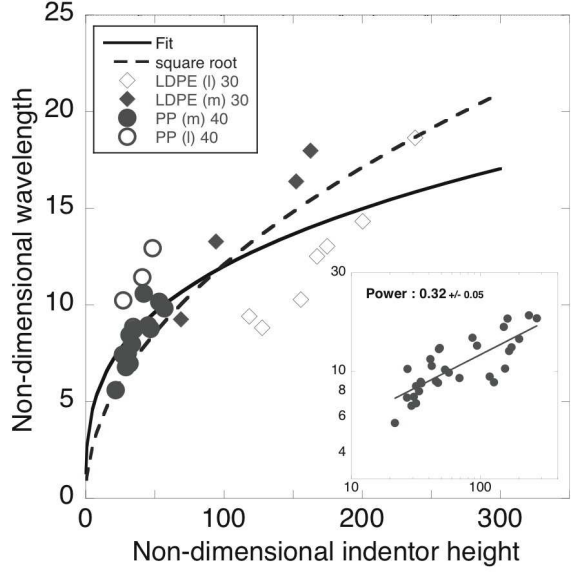}
\caption{The instability. Non-dimensional wavelength $\lambda /[ h( {Eh}/{\dg} )^{{1}/{4}} ]$ as a function of the non-dimensional indentor height $d/h$. The data and symbols of figure \ref{fig:instability} are replotted here. The data collapse on a master curve which can be fitted by a power law (inset and solid curve) with an exponent of $0.32\pm 0.05$. The dashed curve shows a best fit to the predicted square root dependence.}
\label{fig:instabilite_results}
\end{figure}

As seen above, the system is unstable whenever sliding occurs. This accounts for the sensitivity of the experimental threshold to the preparation of the sheet and the thickness of the liquid layer: sliding is facilitated by the lubrication of a thicker layer. 

Following \eqref{lambda}, we rescale the experimental data for the wavelength by plotting $\lambda /[ h( {Eh}/{\dg} )^{{1}/{4}} ]$ as a function of  $d/h$. Figure \ref{fig:instabilite_results} shows a satisfactory collapse of the data onto a master curve. The power-law dependence is found to be $0.32\pm 0.05$, which is weaker than the predicted $1/2$ power. This might be attributed to the approximation $\tilde{R}=R$. In fact, $\tilde{R}$ can be smaller than $R$ when the stress field feels the outer boundary of the sheet --- an effect neglected when we determined the stress function \eqref{airyext}. Note that other mechanisms of wavelength selection ~\cite[such as those studied in][]{mora06,huang07} would lead to no dependence on the Young's modulus and so cannot account for our results.

\subsection{Triangular blisters}
\label{sec:TriangularBlisters}

\subsubsection{Theory}
\label{sec:Theory}

Our earlier analysis of a circular blister was based on neglecting the energy due to bending. This simplification was valid for two reasons: firstly the energetic cost of bending is much smaller than that of stretching and secondly an axisymmetric configuration necessarily induces stretching. When the crack front becomes an open curve, the system has more freedom and thus can avoid expensive stretching. Our analysis for triangular blisters assumes that the sheet does not undergo in-plane stretching and so has a developable conical shape~\cite[see also][]{benamar97,cerda99,cerda05}. In our system, the adhered part of the sheet is free to slide eliminating the possibility of a `hoop stress', which is an additional complexity in previous analyses \cite[see][for example]{cerda05}. For simplicity we derive here the governing equation for the deformed shape of the cone, though it may also be recovered as the small deformation limit of equation (4.17) of \cite{cerda05} in the absence of a constraint, i.e.~ $a=1$.

\begin{figure}
\centering
\includegraphics[height=6cm]{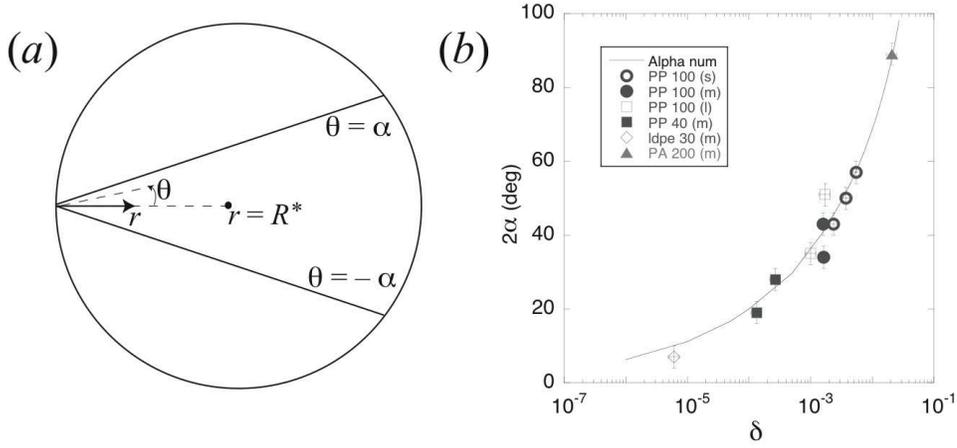}
\caption{The triangular blister. (a) Theoretical idealization of the triangular blisters observed at large displacements. (b) Experimentally measured angle at the tip of the triangle, $2\alpha$, as a function of $\delta = {B d^2}/[\Delta \gamma R{^*} ^4]\, \log{2 R^*}/{r_c}$ where $B$ is the bending stiffness, $\Delta \gamma = 2 \gamma$ is twice the surface tension of the liquid, $R^*$ is the radius of the sheet and $r_c$ is a cut-off which corresponds to the radius of curvature of the tip. We measured $\beta = 4 \alpha$  for three types of plastic material (PP, LDPE, PA) with thicknesses ranging between $30\mu$m for LDPE and $200\mu$m for PA. Three sizes of disk were used: $120$mm (s), $175$mm (m), $275$mm (l). The solid curve shows the theoretical prediction obtained by minimizing \eqref{tri:gdefn}.}
\label{triangle}
\end{figure}

A theoretical idealization of a triangular blister is shown in figure \ref{triangle}a. Here we treat the angle at the tip of the triangle, $2\alpha$, as the unknown parameter to be determined. We use a cylindrical polar co-ordinate system, $(r,\theta)$, centred on the tip of the isoceles triangle. (Note that $r$ is dimensional, since the non-dimensionalization used in earlier sections is now redundant.) The only developable shape compatible with a triangular crack front is a cone, so that
\begin{equation}
w(r,\theta)=d\frac{r}{R^*}f(\theta)
\end{equation} for some unknown function $f(\theta)$. Here, we shall only consider $\theta\geq0$ --- the symmetry of the sheet shows that in fact $f$ depends only on $|\theta|$.

The bending energy of the sheet is then given by
\begin{eqnarray}
{\cal E}_b&\approx&B\int_0^{\alpha}\upd \theta\int_{r_c}^{2R^*}\upd r~r(\nabla^2w)^2\nonumber\\
&=&\frac{Bd^2}{R^{*2}}\log\frac{2R^*}{r_c}\int_0^\alpha(f''+f)^2\upd \theta,
\label{tri:benden}
\end{eqnarray} where $r_c$ is some cut-off length that we assume is unimportant in the following.

Requiring the bending energy in \eqref{tri:benden} to be stationary with respect to variations in $f$, we find that
\begin{equation}
f^{(iv)}+2f^{(ii)}+f=0.
\label{tri:feqn}
\end{equation} Equation \eqref{tri:feqn} has solutions of the form $f(\theta)=(A\theta+B)\sin\theta+(C\theta+D)\cos\theta$ where $A,B,C,D$ are constants to be determined from the boundary conditions, which correspond to continuity of displacement and slope: 
\begin{equation}
f(0)=1,\quad f'(0)=0,\quad f(\alpha)=0,\quad f'(\alpha)=0.
\end{equation} 
Thus the cone profile is given by
\begin{equation}
f(\theta)=\cos\theta+\frac{(\alpha+\sin\alpha\cos\alpha)(\sin|\theta|-|\theta|\cos|\theta|)-\theta\sin\theta\sin^2\alpha}{\alpha^2-\sin^2\alpha}.
\end{equation}

We shall adopt the same approach as earlier writing down the total energy of the system and minimizing this with respect to variations in the angle $\alpha$. The bending energy is given by
\begin{equation}
{\cal E}_b\approx\frac{Bd^2}{{R^*}^2}\frac{2\alpha+\sin2\alpha}{\alpha^2-\sin^2\alpha}\log\frac{2R^*}{r_c}.
\label{tri:bendenf}
\end{equation} Using elementary geometry, the change in surface energy of this configuration relative to the flat state is
\begin{equation}
{\cal E}_s=\dg {R^*}^2(2\alpha+\sin2\alpha),
\label{tri:surfen}
\end{equation} where $R^*$ is the radius of the sheet.

Adding the two energies in \eqref{tri:bendenf} and \eqref{tri:surfen}, we find that the angle $\alpha$ should be chosen to minimize 
\begin{equation}
g(\alpha)\equiv2\alpha+\sin2\alpha+\delta\frac{2\alpha+\sin2\alpha}{\alpha^2-\sin^2\alpha},
\label{tri:gdefn}
\end{equation} where 
\begin{equation}
\delta\equiv\frac{Bd^2}{\dg {R^*}^4}\log\frac{2R^*}{r_c}.
\label{deltadef}
\end{equation}

For minima we require $g'(\alpha)=0$, which can easily be solved numerically, yielding the solid curve shown in figure \ref{triangle}b. For $\alpha\ll1$, the leading order behaviour of $g(\alpha)$ may be used to show that the total included angle at the apex of the triangle (measured in degrees) is
\begin{equation}
2\alpha_c\approx\frac{360}{\pi}(9\delta)^{1/4}.
\label{tri:asyalpha}
\end{equation} 
We note that this scaling is consistent with the scaling analysis of \S 3 (a)i: replacing $R$ with $\alpha R^*$ in \eqref{sqrt:scal} we recover \eqref{tri:asyalpha}. Furthermore, we find  that \eqref{tri:asyalpha} is correct to within  5\% provided that $\delta<0.01$.

\subsubsection{Experimental results}
\label{sec:ExperimentalResults}

In figure \ref{triangle}, we plot the angle at the tip of the triangle as a function of the non-dimensional variable $\delta$  defined by \eqref{deltadef}. We took the cut-off length $r_c$ to be the radius of curvature of the crack front at the tip of the triangle. There is good agreement with the theoretical curve, showing that this regime is ruled by a balance between bending and capillary forces.

\section{Conclusion}
In this article we have considered a thin sheet adhering to a stiff substrate by means of the surface tension of a liquid. We have investigated, experimentally and theoretically, the debonding of the sheet under an imposed central displacement. This is the generic debonding mode whenever the sheet is not peeled from an edge. We uncovered three main regimes dependant on the central displacement. For small vertical displacements,  the debonded area is axisymmetric. This regime was known previously in the context of the conventional blister test \cite[][]{dannenberg61,briscoe91} where there is no liquid. The presence of a liquid in our system allows sliding and so we derived a new analytical solution to account for this. In the second regime, the crack front undulates, and we determined a characteristic wavelength for the undulations. Finally, the crack front becomes triangular, and the shape of the sheet is approximately a developable cone. Overall, our theory is in good agreement with experiments suggesting that the present framework might serve as a reference for the adhesion of elastic sheets.

Our study may also be relevant to small scale biological and technological systems such as the hair of gecko~\cite[][]{autumn02} or microelectromechanical devices~\cite[]{xia99,mastrangelo93b}: capillary and van der Waals forces become increasingly important at small scales. The delamination of thin films in general is slightly different than that studied here. In general settings, debonding is driven by in-plane pre-stress and so there is a compressive azimuthal stress in the debonded part also. However, we expect our result for the wavelength in the undulating regime to hold for the stability of delamination blisters, which apparently has not been investigated previously. Despite this expectation, much more work is required to fully understand the complex patterns generated by the competition between elastic and adhesive forces.

\begin{acknowledgements}
We are grateful to Mokhtar Adda-Bedia for fruitful discussions. D.V. is partially supported by the Royal Commission for the Exhibition of 1851.
\end{acknowledgements}

\appendix{Membrane solution with $\nu\neq1/3$\label{app:vneq13}}

In this Appendix we consider again the membrane solution for the blister given in \S \ref{sec:TheTheory}\ref{sec:circtheory}. Recall that \eqref{mansfieldsol} was used as the solution of the membrane equations but does not satisfy the no-slip boundary condition at the blister edge, \eqref{bcnoslip}, unless $\nu=1/3$. Here, we look for a small perturbation to the approximate solution \eqref{mansfieldsol}  of the form
\begin{equation}
W\approx W_0\left(1-\rho^{2/3}\right)+\delta W(\rho), \quad \Phi\approx \tfrac{3}{16}W_0^2\rho^{4/3}+\delta\Phi(\rho).
\label{perturbset}
\end{equation} Substituting these forms into the membrane equations \eqref{ndfvk2} and \eqref{memlim} and assuming that $\delta W,\delta\Phi\ll1$ we find that
\begin{equation}
W\approx W_0\left(1-\rho^{2/3}+\dn\bigl[\rho^2-\rho^{2/3}\bigr]\right), \; \Phi\approx \tfrac{3}{16}W_0^2\rho^{4/3}\left(1+\dn\bigl[2+\tfrac{3}{2}\rho^{4/3}\bigr]\right),
\label{perturb}
\end{equation} where
\begin{equation}
\dn\equiv \frac{3\nu-1}{17- 15\nu}.
\end{equation} Typically in our experiments $\nu=0.4$, which corresponds to $\dn\approx0.02$ justifying our assumption of a small perturbation. 

Using the value of $\Phi$ in \eqref{perturb} to calculate the stretching energy in the sheet we find, to leading order in $\dn$, that
\begin{eqnarray}
{\cal E}_e\approx \frac{\pi}{16}\left(\tfrac{5}{3}-\nu\right)\frac{Ehd^4}{R^2}\left(1+6\dn\right).
\end{eqnarray} Performing the same total energy minimization as earlier we find that the radius of the blister is
\begin{equation}
R\approx \tfrac{1}{2}\left\{\bigl(\tfrac{5}{3}-\nu\bigr)\frac{Eh}{\dg}\right\}^{1/4}d\left[1+\tfrac{3}{2}\dn+O(\dn^2)\right]
\label{perturbrad}
\end{equation} Using $\nu=0.4$ we find that the correction to the blister radius provided by \eqref{perturbrad} of the earlier expression \eqref{radiusint} is $<3\%$.

\appendix{Analysis of a Two-Dimensional Blister\label{2d:append}}

In this Appendix, we consider the two-dimensional version of the axisymmetric blister test treated in the paper. In particular, we shall determine the effect of a non-negligible bending stiffness on the radius of the blister for a given indentor height and quantify more precisely when it is reasonable to use the membrane approximation. We use a two-dimensional geometry since  it is then possible to make considerable analytical progress. 

For a two-dimensional geometry we find that the Airy stress function is given by $\phi=\sigma_{xx}y^2/2$ for some tension $\sigma_{xx}$. The first F\"{o}ppl-von K\'{a}rm\'{a}n equation then takes the form
\begin{equation}
B\frac{\upd ^4w}{\upd x^4}=\sigma_{xx}\frac{\upd^2w}{\upd x^2}.
\end{equation} 
Using the non-dimensionalizations defined in (3.7) of the paper and letting 
\begin{equation}
\mu^2\equiv \epsilon^{-1}\frac{\sigma_{xx}}{Eh}
\label{2d:alphadefn}
\end{equation}
we find that
\begin{equation}
\frac{\upd^4W}{\upd X^4}=\mu^2\frac{\upd ^2 W}{\upd X^2},
\label{2d:beam}
\end{equation} which is to be solved with the continuity of slope and displacement
\begin{equation}
W(0)=W_0,\quad W'(0)=0,\quad W(1)=W'(1)=0.
\label{2d:beambcs}
\end{equation} The solution of \eqref{2d:beam} with boundary conditions \eqref{2d:beambcs} is
\begin{equation}
W(X)=W_0\frac{(1-|X|)\cosh\tfrac{1}{2}\mu+\mu^{-1}\bigl[\sinh\mu(|X|-\tfrac{1}{2})-\sinh\tfrac{1}{2}\mu\bigr]}{\cosh\tfrac{1}{2}\mu-2\mu^{-1}\sinh\tfrac{1}{2}\mu}.
\end{equation}

Determining the value of $\sigma_{xx}$, and hence $\mu$, requires an additional condition. This condition is the relationship between the strain $\epsilon_{xx}$ and the tension $\sigma_{xx}$~\cite[][]{mansfield}
\begin{equation}
\epsilon_{xx}=\frac{\upd U}{\upd X}+\tfrac{1}{2}\left(\frac{\upd W}{\upd X}\right)^2=\frac{\sigma_{xx}(1-\nu^2)}{Eh}.
\end{equation} Integrating this relationship between $X=0$ and $X=1$ and assuming that the sheet is fixed ($U(0)=U(1)=0$) at the boundaries shows that
\begin{equation}
\frac{2\sigma_{xx}(1-\nu^2)}{Eh}=\int_0^1W'(X)^2~\upd X=W_0^2f_1(\mu),
\label{f1defn}
\end{equation} where
\begin{equation}
f_1(\mu)\equiv \frac{2+\cosh\mu-3\mu^{-1}\sinh\mu}{2\left(\cosh\tfrac{1}{2}\mu-2\mu^{-1}\sinh\tfrac{1}{2}\mu\right)^2}.
\label{f1defn1}
\end{equation}(We note that when a two-dimensional sheet is free to slide, there will be no stretching and only bending of the sheet. This is qualitatively different from the axisymmetric case where, even in the presence of sliding, geometry forces the sheet to stretch.)

For our purposes, it is useful to eliminate $\sigma_{xx}$ from \eqref{f1defn} in favour of $\mu$ by using \eqref{2d:alphadefn}. Introducing
\begin{equation}
\omega\equiv\frac{W_0^2}{2(1-\nu^2)\epsilon}=6\frac{d^2}{h^2},
\label{2d:bdefn}
\end{equation}
we find that $\mu$ satisfies
\begin{equation}
{\mu^2}/{\omega}=f_1(\mu),
\label{alphaeqn}
\end{equation} which has solution $\mu=\mu_c(\omega)$. (The advantage of using  $\omega$ rather than $W_0^2/\epsilon$ is that $\omega$ is independent of the blister radius $R$, which is as yet unknown. $\omega$ measures the vertical displacement relative to the sheet thickness and so is a natural control parameter.)

We continue by considering the limits of small and large $\omega$, which, as we shall see, correspond to small and large values of $\mu$, respectively. From the asymptotic behaviour of $f_1(\mu)$ for small and large $\mu$ we find that
\begin{equation}
\mu_c\approx\omega^{1/2}\times\begin{cases}
\sqrt{6/5},&\omega\ll 1\\
1,&\omega\gg1.
\end{cases}
\end{equation} Note that the two cases $\omega\ll1$ and $\omega\gg1$ give rise to only very slight variations in the asymptotic relations for $\mu_c$.

To determine the blister radius, $R$, we must calculate the deformation energies of the sheet. The bending energy is given by
\begin{equation}
{\cal E}_{b}=B\int_0^Rw''(x)^2~\upd x=\frac{BW_0^2}{R}f_2(\mu);\; f_2(\mu)\equiv \frac{\mu}{2} \frac{\sinh\mu-\mu}{\left(\cosh\tfrac{1}{2}\mu-2\mu^{-1}\sinh\tfrac{1}{2}\mu\right)^2}.
\end{equation} 
The stretching energy is given by
\begin{equation}
{\cal E}_{e}=\sigma_{xx}\int_0^Rw'(x)^2~\upd x=EhR\epsilon W_0^2\mu^2f_1(\mu),
\end{equation} with $f_1(\mu)$ as defined in \eqref{f1defn1}. We can therefore write the total (elastic) energy as
\begin{equation}
{\cal E}/(EhR)= \epsilon W_0^2\left[f_2(\mu_c)+\mu_c^2 f_1(\mu_c)\right]=2\epsilon^2(1-\nu^2)\omega g(\omega)
\end{equation} where 
\begin{equation}
g(\omega)\equiv f_2\bigl[\mu_c(\omega)\bigr]+\mu_c(\omega)^2 f_1\bigl[\mu_c(\omega)\bigr].
\end{equation}
and $\mu_c$ is given by the solution of \eqref{alphaeqn} for a given value of $\omega$. 

Finally, we can now include the surface energy to determine the size of the blister. The surface energy gained is $2\dg R$ so that the total energy of the system is
\begin{equation}
{\cal E}=2\dg R+2(1-\nu^2)EhR\epsilon^2\omega g(\omega).
\end{equation} Recalling that $\epsilon\propto R^{-2}$ we may then minimize this expression by varying $R$. We find that the value of $R$ at which the energy is minimized may be written
\begin{equation}
\frac{R}{d}=\left(\frac{3Eh}{4\dg(1-\nu^2)}\right)^{1/4}\left(\frac{g(\omega)}{\omega}\right)^{1/4}.
\label{2d:radres}
\end{equation} 

\begin{figure}
\centering
\includegraphics[height=7cm]{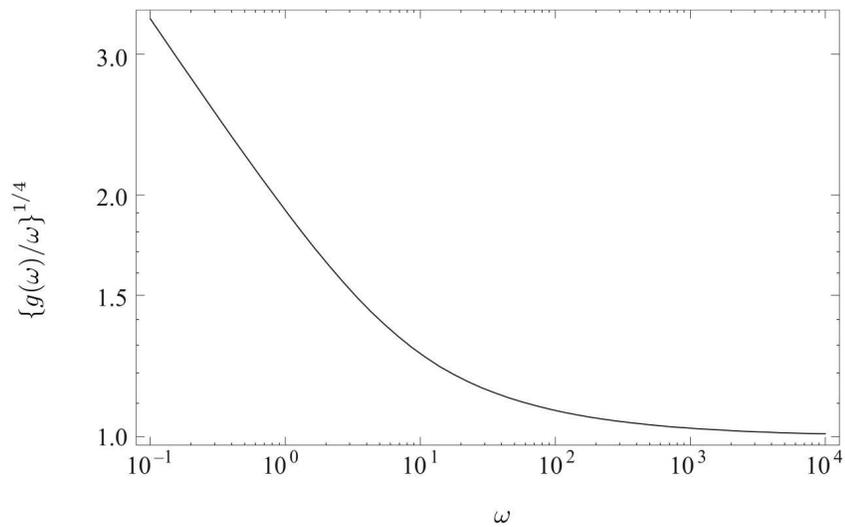}
\caption{The function $\left\{g(\omega)/\omega\right\}^{1/4}$ plotted on logarithmic scales. From \eqref{2d:radres} we see that this function determines the evolution of the blister aspect ratio, $R/d$, as $d$ changes with the material properties of the system fixed. }
\label{2dradius}
\end{figure}

\begin{figure}
\centering
\includegraphics[height=7cm]{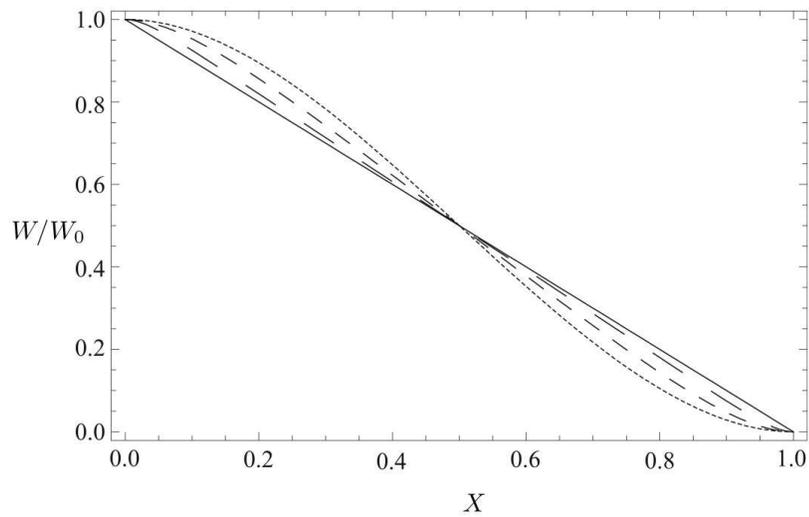}
\caption{Non dimensional blister shapes for different values of the parameter $\omega$, defined in \eqref{2d:bdefn}. Shapes are shown for $\omega=1$ (dotted curve), $\omega=10^2$, $\omega=10^3$ and $\omega=\infty$ (solid line).}
\label{2dshapes}
\end{figure}

The function $\left\{g(\omega)/\omega\right\}^{1/4}$, which determines $R/d$ via \eqref{2d:radres}, is plotted in figure \ref{2dradius}. We note that as $\omega\rightarrow\infty$, $g(\omega)/\omega\rightarrow1$ so that 
\begin{equation}
\frac{R}{d}\approx\left(\frac{3Eh}{4\dg(1-\nu^2)}\right)^{1/4}
\label{2d:largeb}
\end{equation} and we obtain the same scaling relationship as for the axisymmetric analysis. We observe that $\omega>100$ is sufficient for the  relationship \eqref{2d:largeb} to hold to within $10\%$. Furthermore, this corresponds to $d\gtrsim 4h$ --- a condition that is easily satisfied in our experiments. We therefore expect that our neglect of bending effects in the analysis of \S3(a) should be valid except in the earliest stages of the experiment.

For completeness, we note that for $\omega<100$ the prefactor is bigger and indeed that the scaling $R\sim d$ no longer holds. For $\omega\ll1$ we find that $g(\omega)\approx12$ so that
\begin{equation}
\frac{R}{d}\approx\left(\frac{3Eh}{2\dg(1-\nu^2)}\right)^{1/4}\sqrt{\frac{h}{d}}.
\label{2d:smallb}
\end{equation}

\end{document}